\documentclass[%
a4paper,       
10pt, 
UKenglish,     
DIV=10,          
useAMS,usenatbib,referee]%
{scrartcl}

\usepackage[a4paper, total={6in, 9in}]{geometry}
\usepackage{chngcntr}

\usepackage[utf8]{inputenc}
\usepackage[UKenglish]{babel}
\usepackage[T1]{fontenc}
\usepackage{textcomp}
\usepackage{setspace}
\usepackage{hyperref}  
\usepackage{url}                
\usepackage[sumlimits, intlimits, namelimits]{amsmath} 
\usepackage{amssymb}            
\usepackage[caption = true]{subfig} 
\usepackage{array}              
\usepackage{booktabs}           
\usepackage{ae}                 
\usepackage[longnamesfirst,round]{natbib} 
\usepackage{color}              
\usepackage[sc]{mathpazo}       
\usepackage{helvet}             
\usepackage{todonotes}          
\hypersetup{
  bookmarksopen=true, 
  breaklinks=true,
  pdftitle={The sceptical $p$-value},
  pdfsubject={Statistics},
  pdfkeywords={Statistics, Evidence},
  colorlinks=true,
  linkcolor=blue,
  anchorcolor=black,
  citecolor=teal,
  urlcolor=purple
}

\newcommand{\ie}{\textit{i.e.~}}
\newcommand{\pBox}{p_{\mbox{\scriptsize Box}}}
\newcommand{\tBox}{t_{\mbox{\scriptsize Box}}}
\DeclareMathOperator{\Nor}{N} 

\usepackage{doi}

\usepackage{graphicx} 
\usepackage{lscape}
\title{Assessing the replicability of RCTs in RWE emulations}
\author{Jeanette Köppe$^{*}$, Charlotte Micheloud$^{*}$, Stella Erdmann$^{*}$,\\
Rachel Heyard, Leonhard Held \\
{\small $^{*}$ contributed equally}\\
{\small Corresponding author: Jeanette Köppe: jeanette.koeppe@ukmuenster.de}}
\date{\today}
\begin{document}

\maketitle

\begin{abstract}
\section*{Abstract}
\noindent
\textbf{Background}: The standard regulatory approach to assess replication success is the two-trials rule, requiring both the original and the replication study to be significant with effect estimates in the same direction. 
The sceptical $p$-value was recently presented as an alternative method for the statistical assessment of the replicability of study results.\\
\textbf{Methods}: We review the statistical properties of the sceptical $p$-value and compare those to the two-trials rule. We extend the methodology to non-inferiority trials and describe how to invert the sceptical $p$-value to obtain confidence intervals.
We illustrate the performance of the different methods using real-world evidence emulations of randomized, controlled  trials (RCTs) conducted within the RCT DUPLICATE initiative.\\
\textbf{Results}: The sceptical $p$-value depends not only on the two $p$-values, but also on sample size and effect size of the two studies.
It can be calibrated to have the same Type-I error rate as the two-trials rule, but has larger power to detect an existing effect. In the application to the results from the RCT DUPLICATE initiative, the sceptical $p$-value leads to qualitatively similar results than the two-trials rule, but tends to show more evidence for treatment effects compared to the two-trials rule.\\ 
\textbf{Conclusion}: The sceptical $p$-value represents a valid statistical measure to assess the replicability of study results and is especially useful in the context of real-world evidence emulations.\\
\end{abstract}

\textbf{Keywords}: Randomized clinical trial, Real-world evidence, Replication, Sceptical $p$-value, Two-trials rule.

\section{Background}
Randomized controlled trials (RCTs) represent the gold standard for proofing the efficacy of a new therapy \citep{Byar1976, Abel1999, Concato2000}. To reduce bias and guarantee a high internal validity of the results, an RCT is standardized to the highest degree and has strict in- and exclusion criteria. However, older, multi-morbid patients, women (particularly those lactating, pregnant, or able to become pregnant) or non-consenting patients are often excluded from RCTs \citep{Spall2007, FDAWORKSHOP2018} and as a result, a broad generalizability is sometimes missing \citep{Schneeweiss2005, Eichler2011}. Thus, some vulnerable patient groups are under-represented in RCTs, resulting in a lack of evidence for treatment decisions.

Real-world evidence (RWE) plays an increasing role in research and for regulatory decision making, particularly due to an increasing number of potential data sources \citep{FDA2023, Franklin2019, Orsini2020}. RWE is given as the clinical evidence based on so-called real-world data (RWD), i.e. data outside of RCTs given as data relating to patient health status or data routinely collected from different sources \citep{Franklin2021}. With the increase in purely virtual data sources (such as "digital twins" or "Metaverse$^\copyright$"), the term "RWD" is more and more criticized, as all study data come from the "real" world, leading to an increased use of the term "routine practice data" \citep{IQWiG2020}. For the sake of clarity, however, the term RWD will be used as a synonym for routine practice data in the following and includes all type of data outside of RCTs. 

Studies based on RWD can be used for post-market analysis of treatments in actual real-world care to ensure patient safety \citep{Schneeweiss2005, Eichler2011, Franklin2019}, especially for research questions or patients groups that are difficult to address otherwise \citep{Liu2022, Franklin2017, Franklin2019, Booth2019}.
In the literature known as efficacy-effectiveness gap, some studies reported a low performance of drugs and treatment in daily clinical practice \citep{Bhandari2004, Eichler2011}, but systematic analyses failed to proof a general significant difference between the effect estimate in observational studies and in RCTs \citep{Concato2000, Golder2011, Kuss2011, Lonjon2014, Anglemyer2014}.
It is therefore of interest to determine if and when RCTs and well-designed RWE studies, closely aligned with the original RCT, reach the same conclusions. The RCT DUPLICATE initiative (Randomized, Controlled Trials Duplicated Using Prospective Longitudinal Insurance Claims: Applying Techniques of Epidemiology) consists of the emulation of 32 RCTs with RWD to check if and under which circumstances RWD studies explicitly designed to emulate RCTs are useful to draw the same causal conclusions \citep{Franklin2021, Wang2023DUPLICATE}.
A standardized and transparent study protocol for each RWE study was developed, with pre-defined patient selection and definition of primary analyses and study measures. Moreover, all emulation studies were a-priori registered before analyses were done \citep{Franklin2021, Wang2023DUPLICATE}. All emulations were based on data of up to three different US claims data and propensity score matching with 120 pre-defined possible confounders was performed  \citep{Franklin2021, Wang2023DUPLICATE}.

It is common to distinguish direct and conceptual replications \citep{nosek2017}.  A direct replication is an attempt to reproduce a previously observed result based on new data, collected with a protocol as close as possible to the original study. Only the study sample size can be different and is often larger in replication studies to ensure sufficient power to confirm the original finding. In a conceptual replication, the aim is to show the robustness of a finding with a different study sample or study conditions. 
An emulation of an RCT is a conceptual replication, as it tries to confirm the result from an RCT with observational RWD \citep{Wang2017}.
The possible problem of confounding through the lack of randomisation is usually addressed with statistical adjustments such as propensity score matching \citep{Wang2023DUPLICATE}. 

Whether direct or conceptual, the evaluation of the replicability of two studies (e.g. RCT vs replication given by the RWE emulation) is not straightforward and different statistical methods are available \citep{Held2020, Errington2021}. 
The regulatory agreement \citep[also known as the two-trials rule, see e.g.][]{Held2024} is a commonly used method to assess replicability of the two studies \citep{Errington2021, Franklin2021}. Regulatory agreement is fulfilled if the original effect and replication effect go in the same direction and are both significant. However, the pure assessment of the significance of the original trial and the replication study (two-trials rule \textendash~TTR) has the disadvantage that it does not directly take into account the effect sizes of the two studies \citep{Held2022}. 

An alternative approach is to perform a meta-analysis of the RCT and RWE results. The two effect sizes are combined into an overall effect size and then significance is assessed. However, it is important to note that studies entering a meta-analysis are assumed to be interchangeable, which is sometimes a questionable assumption, since the studies were often conducted under different standards. In the case of the replication of an RCT with RWE in particular, this assumption is not fulfilled.

Another approach \textendash~  the so-called “sceptical” $p$-value \textendash~ was recently developed \citep{Held2020, Held2022, Micheloud2023} and represents an alternative to the two-trials rule. It depends on the $p$-values of both studies and also on the variance ratio of both effect sizes. The original trial and the replication study are thus no longer interchangeable, a property which is particularly important with regards to the emulation of RCTs using RWD.

The aim of this paper is to investigate how the sceptical $p$-value performs in this context. 
To do so, the sceptical $p$-value will be used to evaluate the agreement of RWE emulation of RCTs given from the RCT DUPLICATE initiative \citep{Franklin2021, Wang2023DUPLICATE} and the results will be compared to the two-trials rule.
First, we describe the data from the RCT DUPLICATE initiative in Section~\ref{sec:data}. Section~\ref{sec:methods} reviews the sceptical $p$-value and extends it to non-inferiority studies. Type-I error control and power for replication studies are discussed and confidence intervals for the combined effect are derived. In Section~\ref{sec:results}, the sceptical $p$-value is used to evaluate the replicability of emulations of RCTs given by the data set of the RCT DUPLICATE initiative \citep{Franklin2021, Wang2023DUPLICATE} and compared to the two-trials rule. The results are discussed in Section~\ref{sec:discussion}; especially in comparison with the conclusions from the original RCT DUPLICATE initiative.  

\section{Data source: RCT DUPLICATE initiative}\label{sec:data}

For the study at hand, data of 32 RCTs and related RWE emulations from the RCT DUPLICATE initiative were available \citep{Franklin2021, Wang2023DUPLICATE}. Detailed information about the RCT DUPLICATE initiative can be found in the literature \citep{Franklin2020, Franklin2021, Wang2023DUPLICATE, Heyard2024}. In short, published and ongoing RCTs were selected if the trials were relevant to regulatory decision making and were potentially replicable using RWD \citep{Franklin2020, Franklin2021}. RCTs were considered to be potentially replicable if they fulfilled critical aspects concerning study protocol, primary outcome, inclusion and exclusion criteria. RCTs were chosen to get a mix of superiority and non-inferiority trials with large and small magnitudes of effect sizes and trials with active comparators as well as trials with active treatment and placebo added to active standard of care therapies \citep{Franklin2020}. All RCTs were required to be sufficiently large and well powered to guarantee reliable results.

For all emulation studies, the data of up to three different US claims data-sources were used to emulate the related RCTs: \textit{Optum Clinformatics} (from 01/2004 to 03/2019), \textit{IBM Market Scan} (from 2003 to 2017) and a subset of \textit{Medicare Part A, B and D} (2011-2017 including all patients with a diabetes or heart failure diagnosis, 2009-2017 including all patients with a dispensation for an oral anticoagulant). Information about other insurance holders are not available for this database. As a result, 13 of the 32 RCTs were not emulated on the Medicare dataset (TRITON-TIMI, PLATO, ISAR-REACT5, TRANSCEND, ON-TARGET, HORIZON-PIVOTAL, VERO, P04334, D5896, IMPACT, POET-COPD and INSPIRE).
The development of all emulation studies was standardized and transparent, with prospectively defined patient selection, primary analyses and registration \citep{Franklin2021, Wang2023DUPLICATE}. Within a pre-defined time period, all available patient information in the claims data were included to the studies, with inclusion and exclusion criteria as closely as possible adopted from original RCTs. Moreover, a 1:1 nearest-neighbor propensity score matching with caliper width of 0.01 to control for more than 120 possible confounders was performed for all replication studies. Confounders were selected \textit{a priori} and measured during 6 months before drug initiation. Detailed information about confounder selection can be found in \cite{Franklin2021}.

\begin{figure}
    \centering
   \includegraphics[width=1\textwidth]{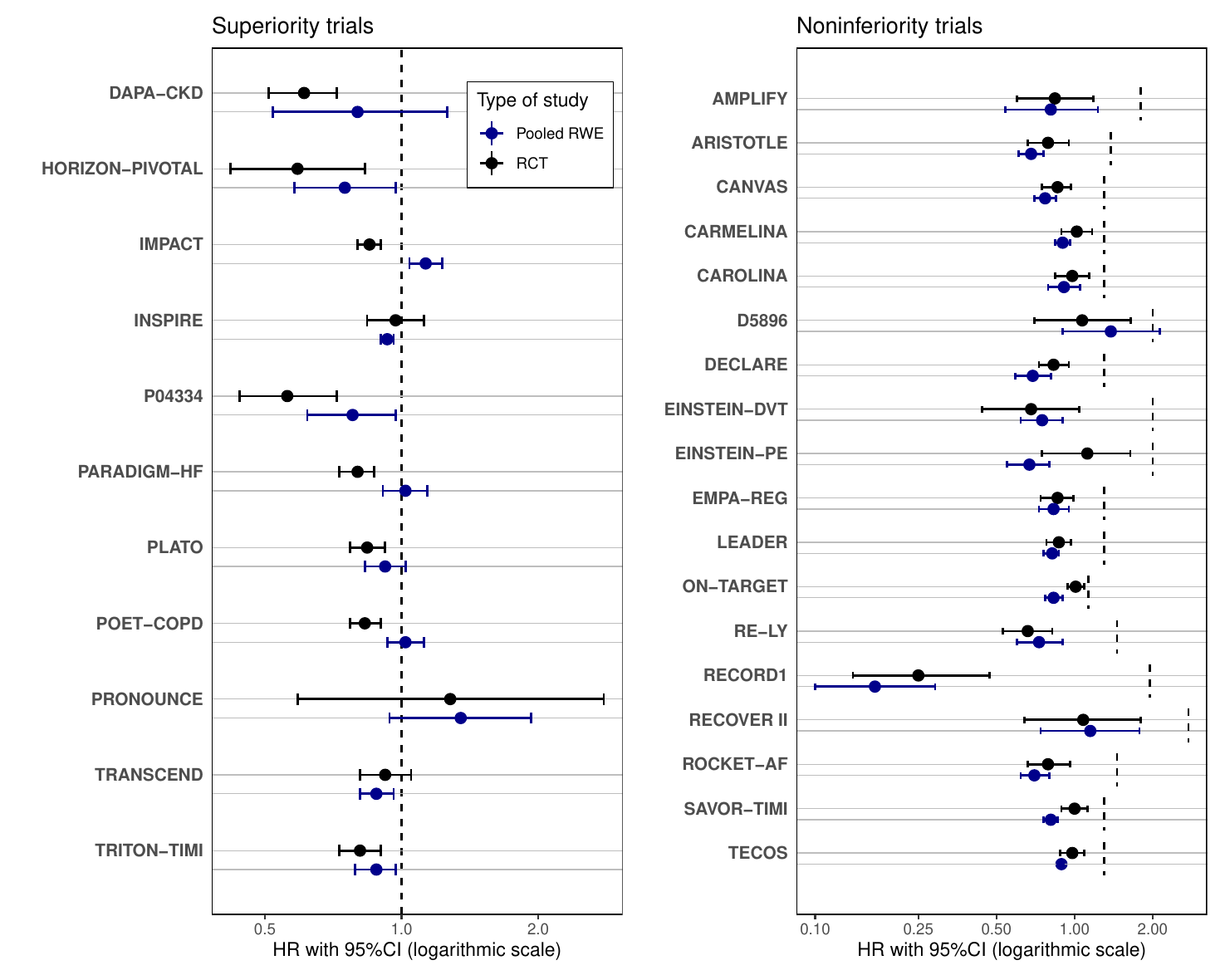}
    \caption{Forest plot comparing RCT and RWE effect estimates (HR with 95\% CI) from the RCT DUPLICATE initiative \citep{Franklin2021, Wang2023DUPLICATE}. Individual non-inferiority margins were added for non-inferiority trials. Three trials were excluded due to a continuous outcome (LEAD2) and significant heterogeneity between the estimates of the three US claims data (ISAR-REACT5 and VERO).}
    \label{fig:forestplot}
\end{figure}

For the analyses presented here, the pooled estimates from fixed effects meta-analysis of up to three different data sources were used for all emulation studies. These pooled estimates have also been analyzed in \citet{Heyard2024} with focus on differences between RCT and RWD effect size estimates due to design and population differences.
For the original trials with a non-inferiority design, the margins were retrieved from the original protocols. For the sake of clarity, only original trials with hazard ratios (HRs) as the primary effect size were included, resulting in the exclusion of the LEAD2 trial due to a continuous outcome. However, in general, the sceptical $p$-value can also be used for continuous or binary outcomes. In addition, two further trials (ISAR-REACT5 and VERO trial) were excluded, because a meta-analysis was not performed due to significant heterogeneity between the estimates of the three US claims data detected by Cochran's $Q$-test. Thus, $N=29$ pairs of original RCTs and related emulation studies (pooled data) were included for further analyses, $11$ with a superiority and $18$ with a non-inferiority design. 
The effect estimates of 28 trials were in the intended direction (HR < 1), out of which 26 were statistically significant. The two non-significant trials (INSPIRE and TRANSCEND) had one-sided $p$-values of $p=0.34$ and $p=0.10$, respectively. One trial (PRONOUNCE) had an effect estimate in the opposite direction (HR > 1), and 
 is one of the two trials (with CAROLINA) for which the emulation was performed  before the results of the RCT were made public. 
Hazard ratios with 95\% confidence interval of all 29 analyzed trials from the RCT DUPLICATE initiative are presented in Fig.~\ref{fig:forestplot} and appendix table \ref{tbl:CIs}.

\section{Methods}\label{sec:methods}
\subsection{The sceptical $p$-value}
The lack of a standard approach to define replication success 
motivated
\citet{Held2020} to develop a new method: the sceptical $p$-value, which combines two principles: `the Analysis of Credibility' by \citet{Matthews2001, Matthews2001b} and a prior-data conflict assessment proposed by \citet{Box1980}.

Assume that the effect estimate from the RCT is in the anticipated direction ($\mbox{HR} < 1$, so $\hat\theta_{\mbox{\scriptsize RCT}} = \log(\mbox{HR}) < 0$) and is significant at level $\alpha$, \ie one-sided $p$-value $p_{\mbox{\scriptsize RCT}} < \alpha$.
This result from the RCT is challenged by a sceptical prior that renders the significant finding no longer convincing (lower limit of the posterior credible interval fixed at zero). The corresponding variance $\tau^2$ of the sceptical prior decreases with increasing (one-sided) $p$-value $p_{\mbox{\scriptsize RCT}}$, so RCTs with a larger $p$-value will lead to a wider sceptical prior. Note that the sceptical $p$-value for the alternative $\mbox{HR} < 1$ is not available, if the effect estimate from the RCT is already in the opposite direction ($\mbox{HR} > 1$), such as in the PRONOUNCE study.

In the second step, the sceptical prior needs to be compared with the RWE effect estimate $\hat \theta_{\mbox{\scriptsize RWE}}$ with standard error $\sigma_{\mbox{\scriptsize RWE}}$. \citet{Held2020} proposed to assess prior-data conflict using the prior-predictive distribution of $\hat\theta_{\mbox{\scriptsize RWE}}$ \citep[Section~5.8]{Spieg2004}.
This results in the test statistic
$\tBox = {\hat\theta_{\mbox{\scriptsize RWE}}}/{\sqrt{\tau^2 + \sigma_{\mbox{\scriptsize RWE}}^2}}$,
which is compared against a standard normal $\Nor(0,1)$ distribution to obtain the tail probability $\pBox = \Phi(\tBox)$ where $\Phi(.)$ denotes the standard normal cdf and smaller values 
of $\pBox$ indicate a larger conflict between the replication effect $\hat \theta_{\mbox{\scriptsize RWE}}$ and the sceptical prior. 
Replication success at level $\alpha$ is achieved if there is significant conflict between the prior and the RWE estimate, i.e.~$\pBox \leq \alpha$. 

The approach needs some adaptions for non-inferiority trials. If $\delta$ denotes the non-inferiority margin on the log hazard ratio scale, then the sceptical prior is now centered at $\delta$. We then obtain the modified test statistic 
\begin{equation}\label{eq:TSmod}
\tBox = ({\hat\theta_{\mbox{\scriptsize RWE}} - \delta})/{\sqrt{\tau^2 + \sigma_{\mbox{\scriptsize RWE}}^2}}.
\end{equation}

\subsubsection{The original formulation}
The value of $\pBox$ depends on the significance level $\alpha$, thus rendering the interpretation difficult. 
The sceptical $p$-value $p_S$ is the smallest level $\alpha$ for which replication success can be achieved, so no longer depends on $\alpha$.
Some algebra shows that the sceptical $p$-value depends on the (one-sided) original and replication $p$-value $p_{\mbox{\scriptsize RCT}}$ and $p_{\mbox{\scriptsize RWE}}$ and the variance ratio $c=\sigma_{\mbox{\scriptsize RCT}}^2/\sigma_{\mbox{\scriptsize RWE}}^2$, here $\sigma_{\mbox{\scriptsize RCT}}$ denotes the standard error of the RCT effect estimate. 
For non-inferiority trials, the $p$-values are calculated for the null hypothesis that the true effect is equal to the non-inferiority margin $\delta$. 

The sceptical $p$-value can be interpreted quantitatively: Smaller $p_S$ values indicate a higher degree of replication success. 
The sceptical $p$-value is one-sided to ensure that replication can only occur if the replication effect estimate has the same sign as the original effect estimate. Further properties of the sceptical $p$-value are studied in \citet{Held2020}. 
First, the approach can be shown to be more stringent than the two-trials rule: if replication success is achieved ($p_S \leq \alpha$), then two-trials rule is also fulfilled ($p_{\mbox{\scriptsize TTR}} = \max\{p_{\mbox{\scriptsize RCT}}, p_{\mbox{\scriptsize RWE}}\} \leq \alpha$)
but not necessarily the other way around. As a consequence, the overall Type-I error rate of the sceptical $p$-value (assuming that both the original and the replication effect estimate is zero) is smaller than the Type-I error rate $\alpha^2$ of the two-trials rule. Second, for fixed original and replication $p$-value, the sceptical $p$-value increases with increasing variance ratio $c$. The variance ratio can be rewritten as the relative sample size $c=n_{\mbox{\scriptsize RWE}} / n_{\mbox{\scriptsize RCT}}$ (replication to original), if we assume that the unit variance in both studies is the same. This implies that a replication study with a large sample size will show a smaller degree of replication success than a replication study with the same replication $p$-value, but a smaller sample size.
The method hence requires replication studies with large sample sizes to be more convincing than those with small sample sizes to achieve the same degree of replication success. 

\subsubsection{Exact Type-I error control}
The sceptical $p$-value has been recently re-calibrated to achieve exact overall Type-I error control across both studies \citep{Micheloud2023} for every value of $c$. The new version has still improved project power compared to the two-trials rule but can now lead to replication success even if one of the two studies is not significant. In what follows we will report the re-calibrated "controlled" version of the sceptical $p$-value.

The exact overall calibration at level $\alpha^2$ implies that we can invert the sceptical $p$-value to obtain one-sided confidence intervals for the combined effect estimate. The upper limit $\theta_u$ of a 97.5\% CI for the log hazard ratio is therefore obtained as the value of $\delta = \theta_u$, where the modified test statistic \eqref{eq:TSmod} gives a squared sceptical $p$-value of 0.025. Exponentiation gives an upper limit on the hazard ratio scale, which can directly be compared to the non-inferiority (or superiority) margin. This approach provides 
an alternative to standard meta-analytic pooling, which may not be compatible to the sceptical $p$-value nor the two-trials rule.

\subsubsection{A comparison}

\begin{figure}
    \centering
    \includegraphics[width=0.9\textwidth]{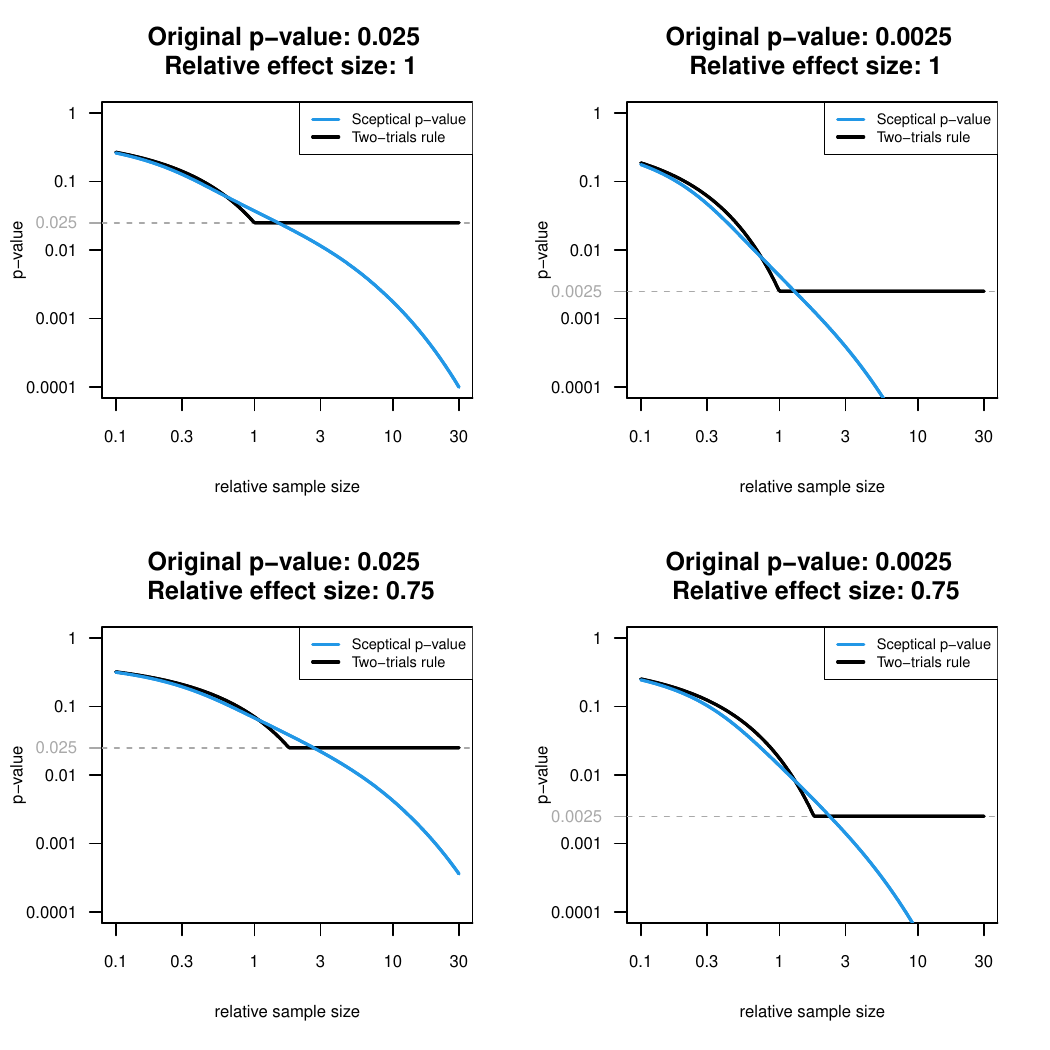}
    \caption{Sceptical $p$-value (square root) and $p$-value from the two-trials rule shown as a function of relative sample size (RWE to RCT) for fixed $p$-value of the original study (dashed grey line) and fixed relative effect size (first row: 1, second row: 0.75).}
    \label{fig:comparison3}
\end{figure}

The (controlled) sceptical $p$-value has exact overall Type-I error control  (across both studies) at level $\alpha^2$ for all values of the variance ratio $c$. The two-trials rule also offers exact Type-I error control at level $\alpha^2$. However, the corresponding $p$-value $p_{\mbox{\scriptsize TTR}}=\max\{p_{\mbox{\scriptsize RCT}},p_{\mbox{\scriptsize RWE}}\}$ cannot be smaller than $p_{\mbox{\scriptsize RCT}}$ nor $p_{\mbox{\scriptsize RWE}}$. 
Figure \ref{fig:comparison3} compares the behaviour of the sceptical $p$-value $p_S$ and the corresponding $p$-value $p_{\mbox{\scriptsize TTR}}$ of the two-trials rule as a function of the relative sample size for fixed original $p$-values.
The calculations assume that the relative effect size $\hat\theta_{\mbox{\scriptsize RWE}}/\hat\theta_{\mbox{\scriptsize RCT}}$ is fixed at 1 (top) and 0.75 (bottom), respectively. Note that for fixed relative effect size, the replication $p$-value depends on the assumed relative effect size and the relative sample size. 

Figure  \ref{fig:comparison3} shows that the sceptical $p$-value behaves very similar to the two-trials rule if the relative sample size is small. However, it decreases smoothly towards zero and can get smaller than $p_{\mbox{\scriptsize RCT}}$ if the relative sample size $c$ becomes large, whereas the two-trials rule $p$-value has an edge and then stays constant, as it cannot get smaller than $p_{\mbox{\scriptsize RCT}}$. 

\subsection{Power of the replication study}
\begin{figure}
    \centering
   \includegraphics[width = \textwidth]{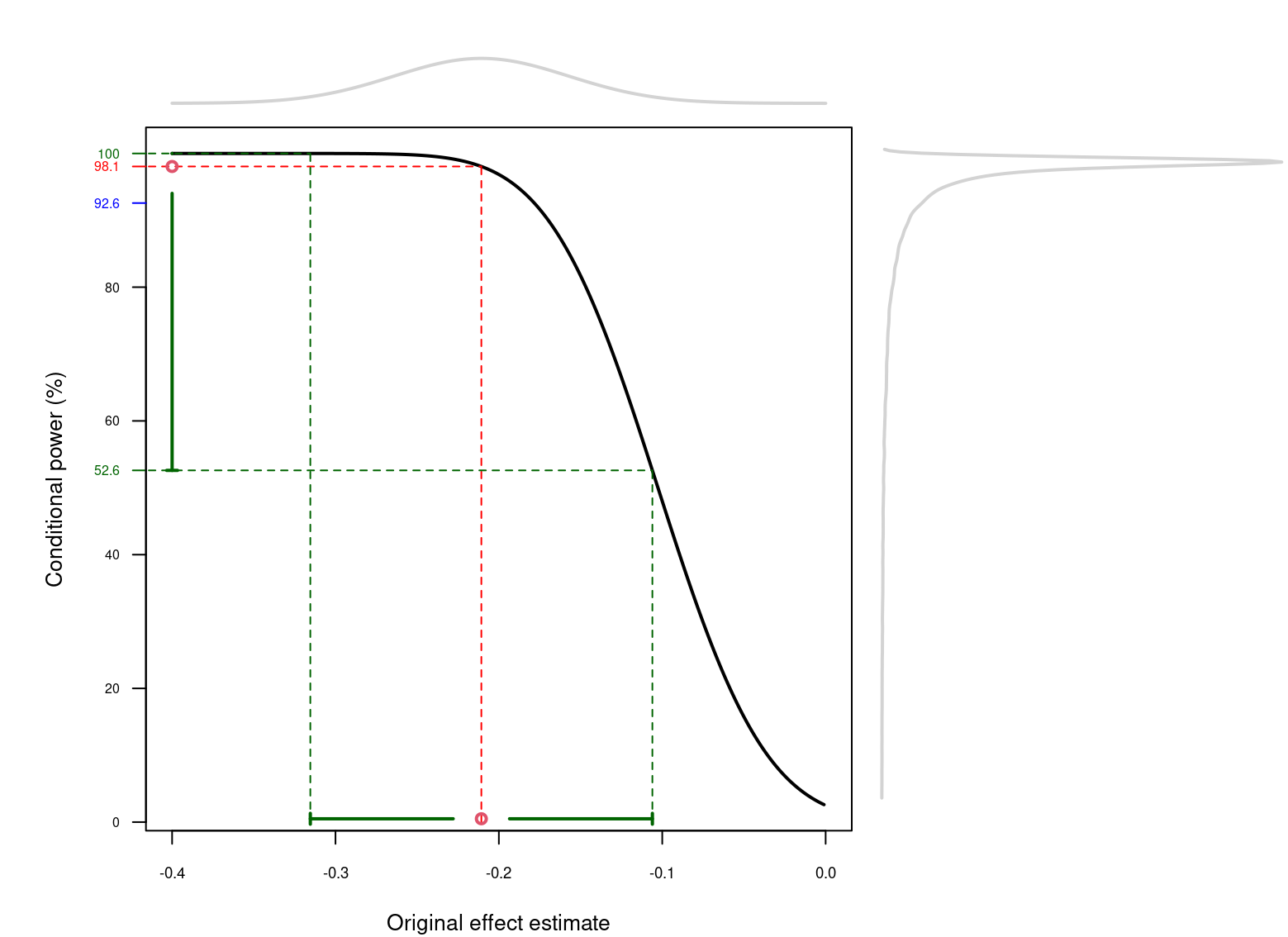}
    \caption{Conditional power as a function of the original effect estimate. The original effect estimate $\hat\theta_{\mbox{\scriptsize RCT}} = -0.21$ of the TRITON-TIMI trial
    and the corresponding conditional power of 98.1\% are colored in red. The grey lines in 
    the top and right axis are the distributions of the original effect estimate and 
    the corresponding conditional power, respectively. The average conditional power of 92.6\% (in blue) is the predictive power. The green intervals represent the range given by the 95\% confidence interval of the original effect estimate.}
    \label{fig:pow}
\end{figure}

The power of the replication study, \emph{i.e.} the probability that a replication study with a certain sample size leads to replication success, has been extensively studied \citep[see for example][]{Anderson2017, Anderson2022, vanZwet2022, Micheloud2022, Micheloud2023}. 
The power is usually calculated as the probability to reach a significant replication result \emph{conditional} on the effect estimate from the original trial.
This concept is illustrated in Figure~\ref{fig:pow} using the TRITON-TIMI trial from the RCT DUPLICATE initiative. The conditional power of the two-trials rule at the one-sided level $\alpha = 0.025$ is plotted as a function of the effect estimate from the RCT (log hazard ratio) assuming that the variance is the same in the original trial and the emulation study ($c = 1$). The original effect estimate in the TRITON-TIMI example is  $\hat\theta_{\mbox{\scriptsize RCT}} = -0.21$, so with $c = 1$, a conditional power of 98.1\% is achieved. However, the original effect estimate comes with uncertainty, represented by the density in the top axis, and the 95\% confidence interval $[-0.31, -0.11]$ around the original effect estimate. The conditional power for the values inside this confidence interval greatly vary, from 52.6\% to 100\%.  One way to incorporate this uncertainty is to compute the \emph{predictive} power \citep{SpiegFreed1986} by averaging the conditional power with respect to the distribution of the original effect estimate. The density of the conditional power is shown in the right axis of Figure~\ref{fig:pow} and the average is 92.6\%, so considerably smaller than 98.1\%. 

If the power to detect the effect from the original study is not too large (< 95\%), the conditional Type-I error rate (the probability of replication success if the replication effect estimate is zero) of the sceptical $p$-value is shown to be bounded and never larger than $2 \alpha$ \citep[Section 3.4]{Micheloud2023}. Such a two-fold increase in Type-I error rate is the price to be paid for increased power, but considered to be acceptable even in the regulatory setting \citep{Rosenkrantz2002}. 

Power calculations are also used to determine which sample size is required in the replication study to reach a certain power, usually between 80\% and 90\%. While replication sample size calculation is not relevant in the RCT DUPLICATE study, where all available patient information in the claims data was used, it is of interest to know the power of a RWE study with the available data. It is therefore advisable to determine the power for an existing dataset before calculating the effect estimates for the emulation, to ensure that the data are indeed sufficient. Conversely, calculating the minimum necessary sample size can also be beneficial in the context of RWE, for example, to understand the cohort size required after propensity score matching. If the power is insufficient or if the sample size is inadequate after matching, the chosen dataset is likely not suitable for the planned emulation. In Section~\ref{sec:results}, we calculate the conditional and predictive power of the two-trials rule and of the sceptical $p$-value for all the RWE emulations \citep{Micheloud2022, Micheloud2023}. 

\section{Results}\label{sec:results}

\subsection{Replication of RCTs}
Results of the replication success assessment with the two methods are presented in Table \ref{tab:pvalues_RWE}. The variance ratio $c$ is larger than 10 for two studies (TECOS, INSPIRE), see also Figure \ref{fig:forestplot}. TECOS is a non-inferiority study, where both the RCT and the RWE study showed non-inferiority with respect to a HR margin of 1.4. INSPIRE is a superiority study, where both effect estimates have been close to 1, but the RWE estimate is slightly smaller and more much precise. The sceptical $p$-value is smaller or equal to the $p$-value from the two-trials rule for the vast majority of the studies (27/29 studies). Note also that in three cases (CARMELINA, TRANSCEND, INSPIRE) the sceptical $p$-value is smaller than $p_{\mbox{\scriptsize RCT}}$ and in seven cases (TRITON-TIMI, PLATO, HORIZON-PIVOTAL, DAPA-CKD, P04334, D5896, and PRONOUNCE) it is smaller than $p_{\mbox{\scriptsize RWE}}$, both is impossible for $p_{\mbox{\scriptsize TTR}}$. 

At the one-sided level $\alpha = 0.025$, the same conclusion is drawn with the two-trials rule and the sceptical $p$-value: The same
20 out of 29 (69\%) emulations successfully replicate the corresponding original RCT results. However, it was noticeable that more studies failed to replicate the original effect, if no Medicare data were available for the pooled effect. If Medicare data were available, 16 of 19 (84\%) emulations successfully replicated the original effect, compared to the lower rate (5 of 10, 50\%) of successfully replications, when Medicare data was not available. This finding was independent from the used metric, i.e. the sceptical $p$-value and the $p$-value from the two-trials rule lead to the same rates of success. 

\begin{table}[ht]
\centering
\begin{tabular}{rlllrll}
  \hline
 & Study & $p_{\mbox{\scriptsize RCT}}$ & $p_{\mbox{\scriptsize RWE}}$ & $c$ & $p_{\mbox{\scriptsize TTR}}$ & $p_S$ \\ 
  \hline
1 & LEADER & < 0.0001 & < 0.0001 & 2.6 & < 0.0001 & < 0.0001 \\ 
  2 & DECLARE & < 0.0001 & < 0.0001 & 0.7 & < 0.0001 & < 0.0001 \\ 
  3 & EMPA-REG & < 0.0001 & < 0.0001 & 1.2 & < 0.0001 & < 0.0001 \\ 
  4 & CANVAS & < 0.0001 & < 0.0001 & 1.8 & < 0.0001 & < 0.0001 \\ 
  5 & CARMELINA & 0.0003 & < 0.0001 & 4.2 & 0.0003 & < 0.0001 \\ 
  6 & TECOS & < 0.0001 & < 0.0001 & 14.3 & < 0.0001 & < 0.0001 \\ 
  7 & SAVOR-TIMI & < 0.0001 & < 0.0001 & 3.5 & < 0.0001 & < 0.0001 \\ 
  8 & TRITON-TIMI & < 0.0001 & 0.007 & 1.0 & 0.007 & 0.003 \\ 
  9 & PLATO & < 0.0001 & 0.056 & 0.7 & 0.056 & 0.031 \\ 
  10 & ARISTOTLE & < 0.0001 & < 0.0001 & 2.7 & < 0.0001 & < 0.0001 \\ 
  11 & RE-LY & < 0.0001 & < 0.0001 & 1.2 & < 0.0001 & < 0.0001 \\ 
  12 & ROCKET-AF & < 0.0001 & < 0.0001 & 2.2 & < 0.0001 & < 0.0001 \\ 
  13 & EINSTEIN-DVT & < 0.0001 & < 0.0001 & 5.3 & < 0.0001 & < 0.0001 \\ 
  14 & EINSTEIN-PE & 0.0018 & < 0.0001 & 4.4 & 0.0018 & < 0.0001 \\ 
  15 & RECOVER II & 0.0002 & < 0.0001 & 1.4 & 0.0002 & 0.0002 \\ 
  16 & AMPLIFY & < 0.0001 & < 0.0001 & 0.7 & < 0.0001 & < 0.0001 \\ 
  17 & RECORD1 & < 0.0001 & < 0.0001 & 1.3 & < 0.0001 & < 0.0001 \\ 
  18 & TRANSCEND & 0.10 & 0.002 & 2.3 & 0.10 & 0.064 \\ 
  19 & ON-TARGET & 0.001 & < 0.0001 & 0.9 & 0.001 & < 0.0001 \\ 
  20 & HORIZON-PIVOTAL & 0.001 & 0.014 & 1.8 & 0.014 & 0.01 \\ 
  21 & DAPA-CKD & < 0.0001 & 0.16 & 0.2 & 0.16 & 0.15 \\ 
  22 & PARADIGM-HF & < 0.0001 & 0.63 & 0.6 & 0.63 & 0.65 \\ 
  23 & P04334 & < 0.0001 & 0.015 & 1.2 & 0.015 & 0.004 \\ 
  24 & D5896 & 0.002 & 0.046 & 1.0 & 0.046 & 0.03 \\ 
  25 & IMPACT & < 0.0001 & 1.00 & 0.5 & 1.00 & 1.00 \\ 
  26 & POET-COPD & < 0.0001 & 0.66 & 0.7 & 0.66 & 0.68 \\ 
  27 & INSPIRE & 0.34 & < 0.0001 & 19.9 & 0.34 & 0.25 \\ 
  28 & CAROLINA & 0.0001 & < 0.0001 & 1.2 & 0.0001 & < 0.0001 \\ 
  29 & PRONOUNCE & 0.73 & 0.95 & 4.7 & 0.95 & NA \\ 
   \hline
\end{tabular}
    \caption{One-sided $p$-values $p_{RCT}$ of the original trial and 
    $p_{RWE}$ of the emulation, as well as variance ratio $c$ and 
    one-sided $p$-values $p_{\mbox{\scriptsize TTR}}$ of the two-trials rule 
    and $p_S$ of the sceptical $p$-value.}
    \label{tab:pvalues_RWE}
\end{table}

\begin{figure}
    \centering
   \includegraphics[width = \textwidth]{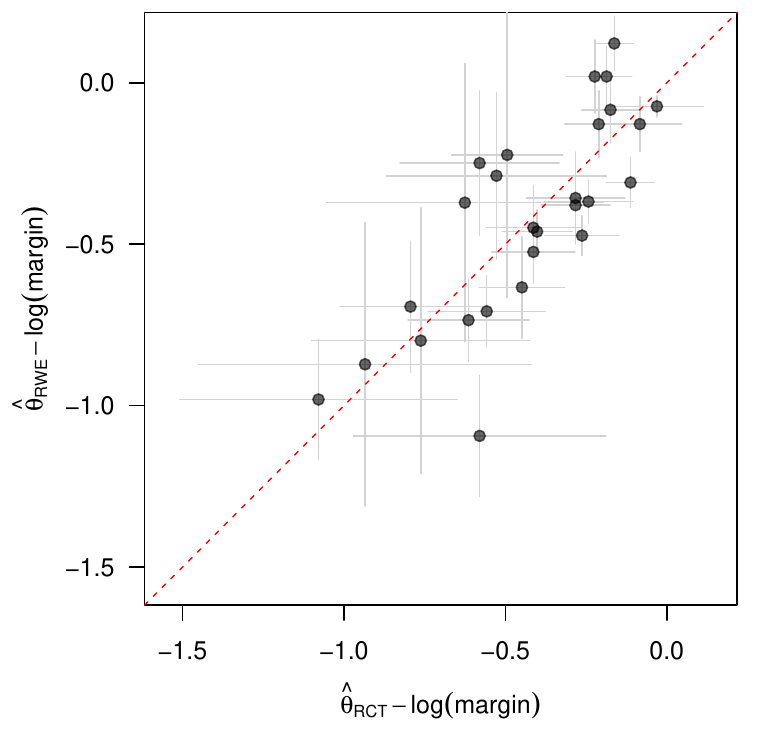}
    \caption{$\hat\theta_{\mbox{\small RCT}}$ - log(margin) versus 
$\hat\theta_{\mbox{\small RWE}}$ - log(margin) with their respective 95\%
confidence interval for the studies 
in the RCT DUPLICATE initiative (without RECORD1, where differences are larger).}
    \label{fig:shrinkage}
\end{figure}

Figure~\ref{fig:shrinkage} shows $\hat\theta_{\mbox{\small RCT}}$ - log(margin) versus $\hat\theta_{\mbox{\small RWE}}$ - log(margin) for the 29 study pairs in the RCT DUPLICATE initiative. The margin is $1$ in the superiority studies. Points below the diagonal represent cases where the effect estimate from the RWE study is shrunken as compared to the effect estimate from the original trial. This shrinkage is towards $0$ for superiority studies and towards the margin for non-inferiority studies. Replication effect estimates usually present a large amount of shrinkage as compared to their original counterpart in recent large-scale replication projects \citep[e.g. in][]{OSC2015, Camerer2016, Camerer2018}. This is not the case here. 

\subsection{Replication power}

Results are displayed in Table~\ref{tbl:condpredPower}. The average conditional power \citep{Amrhein2019, McShane2020} is 85.9\% with the two-trials rule and 87.0\% with the sceptical $p$-value. These values reduce to 83.5\%, and 85.0\%, respectively, for predictive power. If the true underlying effect is the same in the original trial and in the emulation study and both were conducted with high standards, we would expect the average predictive power to be equal to the proportion of success, which is smaller in the RCT DUPLICATE study (69\% with both methods). The proportion of replication success as well as the average predictive power in the RCT DUPLICATE initiative for other values of $\alpha$ are shown in Figure~\ref{fig:propSuccess}.

The three RCTs with a non-significant original $p$-value (PRONOUNCE, INSPIRE and TRANSCEND) have a power of $0$ with all three methods. 
By definition, the power based on a non-significant original trial is always $0$ with the two-trials rule. There is no such requirement on $p_{\mbox{\scriptsize RCT}}$ for the sceptical $p$-value but large relative sample sizes are required to reach an acceptable power level for non-significant original studies.

For the 26 other studies, the power of the two-trials rule and the sceptical $p$-values is very similar, with slightly higher values with the sceptical $p$-values. This is because the original $p$-values are relatively small in every case, which is favored by the sceptical $p$-values 
\citep[see][Sections~3.1 and ~3.4, respectively]{Held2022, Micheloud2023}.

\subsection{Confidence intervals}
We have also calculated confidence intervals for the hazard ratio based on inversion of the sceptical $p$-value and compared those to the ones obtained from applying a fixed effect meta-analysis to the results from the RCT and RWD. The results are given in Table \ref{tbl:CIs} in Appendix \ref{sec:CIApp}. The upper limit of both confidence intervals are similar, the upper limit of the sceptical CI being often slightly larger than the meta-analytic one. The differences get larger when the uncertainty of the effect estimates is substantially different, for example for DAPA-CKD or INSPIRE, where the meta-analytic pooled effect estimate is dominated by the larger study. This is also the case for the three superiority studies PARADIGM-HF, IMPACT and POET-COPD, where the meta-analytic confidence interval indicates significance (at the standard two-sided 95\% level) with an upper CI limit below 1, although the RWE effect estimate is above 1 in all three cases. In contrast, the upper limit of the sceptical CI is always above one, in line with the large sceptical $p$-values for those studies reported Table \ref{tab:pvalues_RWE}. This illustrates that meta-analytic pooling can still flag a positive (significant) finding, even if one of the two studies has an effect in the wrong direction \citep{Held2024}.

\begin{table}[ht]
\centering
\begin{tabular}{rllrrrr}
  \hline
    & && \multicolumn{2}{c}{Conditional power} & \multicolumn{2}{c}{Predictive power}   \\ \cmidrule{4-5} \cmidrule{6-7}
 & Study & $p_{\mbox{\scriptsize RCT}}$ & two-trials rule & sceptical $p$ & two-trials rule & sceptical $p$ \\ 
  \hline
1 & LEADER & < 0.0001 & 100.0 & 100.0 & 100.0 & 100.0 \\ 
  2 & DECLARE & < 0.0001 & 100.0 & 100.0 & 99.7 & 99.9 \\ 
  3 & EMPA-REG & < 0.0001 & 100.0 & 100.0 & 99.8 & 99.9 \\ 
  4 & CANVAS & < 0.0001 & 100.0 & 100.0 & 100.0 & 100.0 \\ 
  5 & CARMELINA & 0.0003 & 100.0 & 100.0 & 98.8 & 99.3 \\ 
  6 & TECOS & < 0.0001 & 100.0 & 100.0 & 100.0 & 100.0 \\ 
  7 & SAVOR-TIMI & < 0.0001 & 100.0 & 100.0 & 99.9 & 99.9 \\ 
  8 & TRITON-TIMI & < 0.0001 & 98.1 & 99.2 & 92.6 & 95.3 \\ 
  9 & PLATO & < 0.0001 & 91.2 & 95.0 & 84.8 & 89.3 \\ 
  10 & ARISTOTLE & < 0.0001 & 100.0 & 100.0 & 100.0 & 100.0 \\ 
  11 & RE-LY & < 0.0001 & 100.0 & 100.0 & 100.0 & 100.0 \\ 
  12 & ROCKET-AF & < 0.0001 & 100.0 & 100.0 & 100.0 & 100.0 \\ 
  13 & EINSTEIN-DVT & < 0.0001 & 100.0 & 100.0 & 100.0 & 100.0 \\ 
  14 & EINSTEIN-PE & 0.0018 & 100.0 & 100.0 & 100.0 & 97.0  \\ 
  15 & RECOVER II & 0.0002 & 98.7 & 99.4 & 92.4 & 94.9 \\ 
  16 & AMPLIFY & < 0.0001 & 95.2 & 97.6 & 90.1 & 93.7 \\ 
  17 & RECORD1 & < 0.0001 & 100.0 & 100.0 & 100.0 & 100.0 \\ 
  18 & TRANSCEND & 0.10 & 0.0 & 0.0 & 0.0 & 0.0 \\ 
  19 & ON-TARGET & 0.001 & 80.5 & 85.6 & 73.4 & 77.9 \\ 
  20 & HORIZON-PIVOTAL & 0.001 & 98.0 & 98.9 & 89.3 & 91.8 \\ 
  21 & DAPA-CKD & < 0.0001 & 59.1 & 65.6 & 58.5 & 64.6 \\ 
  22 & PARADIGM-HF & < 0.0001 & 97.3 & 98.7 & 93.5 & 96.1 \\ 
  23 & P04334 & < 0.0001 & 99.9 & 100.0 & 98.2 & 99.1 \\ 
  24 & D5896 & 0.002 & 81.2 & 85.7 & 73.5 & 77.5 \\ 
  25 & IMPACT & < 0.0001 & 96.7 & 98.3 & 93.4 & 95.9 \\ 
  26 & POET-COPD & < 0.0001 & 97.6 & 98.9 & 93.4 & 96.0 \\ 
  27 & INSPIRE & 0.34 & 0.0 & 0.0 & 0.0 & 0.0 \\ 
  28 & CAROLINA & 0.0001 & 97.3 & 98.8 & 90.6 & 93.7 \\ 
  29 & PRONOUNCE & 0.73 & 0.0 & 0.0 & 0.0 & 0.0 \\ 
  30 & Average &  & 85.9 & 87.0 & 83.5 & 85.0 \\ 
   \hline
\end{tabular}
\caption{Conditional and predictive power (in \%) at the one-sided level $\alpha = 0.025$ with the two-trials rule and the sceptical $p$-value.} 
\label{tbl:condpredPower}
\end{table}

\section{Discussion}\label{sec:discussion}
A target trial emulation is a conceptual emulation of a "hypothetical" RCT, using a two-step process \citep{Hernan2016, Hernan2022}. In the first step, the causal questions are defined in a protocol for the hypothetical RCT, which has to specify the key elements to define the causal estimands (i.e. eligibility criteria, treatment strategies, treatment assignment, the start and end of follow-up, outcomes, causal contrasts) and the data analysis plan \citep{Hernan2016}. The RCT described in the protocol is defined as the target study for causal inference, which will be emulated using observational data in the second step \citep{Hernan2016, Hernan2022}. Furthermore, several studies used an existing RCT instead of a hypothetical RCT, which will be emulated by the target trial emulation \citep{Denge2022, Merola2022, Leening2024, Walker2024, Petito2020,McGrath2021, Khosrow-Khavar2022, Gallivan2023, Yoon2023}. However, even with the existing RCT as a template for the target trial emulation, the effect estimates of the RWD emulation are not always compared with those of the original trial \citep{Denge2022, Petito2020, McGrath2021, Khosrow-Khavar2022, Gallivan2023}.
A statistical comparison of the emulation with the original RCT further substantiates the results and increase the resulting evidence. As seen in the data presented, the sceptical $p$-value can be an excellent extension to evaluate the agreement and thus complements the evidence from target trial emulations of existing RCTs.

The emulations of the 32 evaluated RCTs were previous implemented and analyzed by the RCT DUPLICATE Initiative \citep{Wang2023DUPLICATE}. In this project, \cite{Wang2023DUPLICATE} used different binary agreement metrics to check the replication success of the 32 emulations using claims data of three different US insurance companies. First, named \textit{full statistical significance agreement}, is fulfilled, if the estimates (log(HR)) and the related 95\% CIs are on the same side of the null (without consideration of the predefined non-inferiority margins in the case of non-inferiority trials). 23 (72\%) of the 32 emulations met statistical significance agreement \citep{Wang2023DUPLICATE}. This binary agreement metric should be equivalent to the two-trials rule. However, for two studies, i.e. D5896 and PRONOUNCE, the $p$-value of the two-trials rule  delivered a different conclusion of the replication success. \cite{Wang2023DUPLICATE} reported that statistical agreement was fulfilled for all of the two emulations, whereby in our analysis, $p_{\mbox{\scriptsize TTR}}> 0.025$ holds. In the case of PRONOUNCE, the original RCT had a non-significant treatment effect, resulting in an non-significant $p_{\mbox{\scriptsize TTR}}$ due to the definition of the $p$-value. Moreover, the formalism of $p_{\mbox{\scriptsize TTR}}$ is not meaningful in the case of non-significant effects of the original trial. In the case of the D5896 trial, \cite{Wang2023DUPLICATE} did not use the non-inferiority margins for the definition of full statistical agreement, as it was simply defined as having point estimates and confidence intervals on the same side of null. Using sceptical $p$-value, it is possible to account for the non-inferiority margins, which is one major advantage of this formalism compared to the binary agreement metrics.  \\
As the second binary agreement metric, \cite{Wang2023DUPLICATE} analysed estimated agreement, whereby the estimate of the emulation has to be an element of the 95\% CI of the original effect, which was met for 66\% of the emulations. As the last one, \cite{Wang2023DUPLICATE} evaluated, if the absolute standardized differences of the estimates are smaller then 1.96, which was fulfilled in 72\% of the emulations.
Using the sceptical $p$-value, 69\% of the emulations were evaluated as successful replications. Beside the emulations with non-significant original effects, the findings of the study at hand are in accordance to findings of \cite{Wang2023DUPLICATE}.

However, in general, it is not always possible to replicate the results of an RCT, especially if there are fundamental design differences between the RCT and RWE \citep{Wang2023DUPLICATE, Heyard2024}. \cite{Wang2023DUPLICATE} stated that the conceptual emulation of the original RCT leads to an independent study that emulate the design of the original trial instead of replicating the population of the RCT. Potential differences of the two population (e.g. different distribution of comorbidities) may result in differences in the effect size estimates \citep{Wang2023DUPLICATE}. Moreover, \cite{Heyard2024} analyzed differences between the original RCTs and the related RWE emulations using meta-regression models. They concluded that most of the heterogeneity between RCT and RWD emulation could be explained by delayed effect of treatment, discontinuation of treatment during run-in period, and treatment started in hospital \citep{Heyard2024}. The sceptical $p$-value itself cannot distinguish between replication failure due to design differences, incorrect model specification or differences due to population differences. Instead, the sceptical $p$-value evaluates and quantifies the agreement between two effect estimators. A successful replication of an RCT using RWD, indicated by a significant sceptical $p$-value, suggests the external validity of the RCT results. Conversely, if the replication is unsuccessful, the p-value does not provide insight into the reasons. These issues must be addressed using other methods. In studies based on RWD, it is essential to adequately address data heterogeneity methodologically, for example, using propensity score methods. An important aspect is achieving sufficient balance between the treatment groups. However, it is possible that the treatment effect observed in an RCT cannot be replicated in RWD due to fundamentally different cohorts. This is less a failure of RWE and may indicate that external validity is limited.

There are other studies available that emulate RCTs using RWD, where the binary metrics regulatory agreement, estimated agreement and standardized difference agreement were often used to evaluate the agreement of the effect estimates from the original trials and the related emulation study \citep{Jang2022, Jin2022, Merola2023, Antoine2024, Signori2024, Elvira2024}. However, the two-trials rule and the sceptical $p$-values are good measures to evaluate the agreement between an original RCT and the corresponding emulation study and have benefits compared to binary metrics. Besides the advantages discussed above, both methods allow to calculate the power of the emulation study based on the effect estimate found in the RCT, with and without 
uncertainty incorporation. For all 29 studies considered here, the conditional and predictive power is larger with the sceptical $p$-values than with the two-trials rule. Furthermore, the replication rate is in general slightly higher with the sceptical $p$-value.

In this paper, we focused on the controlled version of the sceptical $p$-value, to ensure a fair comparison with the two-trials rule in terms of overall Type-I error control. An alternative to the controlled re-calibration is the golden re-calibration of the sceptical $p$-value \citep{Held2022}. The golden version is less stringent than the original formulation, because, to establish replication success, original trial and replication study do not both necessarily need to be significant at level $\alpha$. However, a borderline significant original trial ($p_{\mbox{\scriptsize RCT}} = \alpha$) cannot lead to replication success, if there is shrinkage of the replication effect estimate. As shrinkage of the effect estimate from the emulation study as compared to effect estimate from the original trial seems to not be an issue in this context, one could favor the controlled sceptical $p$-value over the golden sceptical $p$-value. Moreover, a comparison of the golden version with the two-trials rule is more difficult, as the two methods have different Type-I error rates. 

The presented work had several limitations. No formalism to assess the replication success of an original null effect using the sceptical $p$-value was available \citep{Pawel2024}. Recently \cite{MicheloudHeld2023} provided an extension of the sceptical $p$-value to equivalence studies. However, a trial that was initially planned for superiority, but fails to show a significant treatment effect, should not be interpreted as an equivalence study and, thus, the formalism for equivalence studies should not be assigned in this case. Instead of performing a second study with the wrong design, a new study with the aim of proofing the equivalence should be initialized (and then replicated). Moreover, as stated above, fundamental design differences between RCT and RWD emulation may result in bias, which will also affect the effect estimate of the RWD study.
\section{Conclusion}
The sceptical $p$-value represents a valid statistical measure to evaluate the replicability of the study results, especially useful in the context of real-world evidence.
\section{List of abbreviations}
\begin{table*}[!h]
\begin{tabular*}{\columnwidth}{ll}
\toprule
Abbreviation & Explanation\\
\midrule
CI	        & Confidence interval\\
HR          & Hazard ratio\\
RCT	        & Randomized controlled trial\\
RWD	        & Real world data\\
RWE	        & Real world evidence\\
TTR         & Two-trials rule\\
\bottomrule
\end{tabular*}
\end{table*}

\section{\textbf{Declarations}}
\subsection{Ethics approval and consent to participate}
Not applicable.
\subsection{Consent for publication}
Not applicable.
\subsection{Availability of data and materials}
The R package \texttt{ReplicationSuccess}, available on CRAN at
\url{https://CRAN.R-project.org/package=ReplicationSuccess}, has been
used for the computation of the sceptical $p$-value and all power calculations. The data and the code to reproduce all the Figures and Tables can be found at \url{https://github.com/CharlotteMicheloud/RWE}.
\subsection{Competing interests}
JK reports research funding from the German Society for Trauma Surgery sponsored by Stryker, outside the submitted work. 
SE, CM, RH and LH have no conflict of interest to declare. 
\subsection{Funding}
None.
\subsection{Authors' contributions}
\textbf{Jeanette Köppe:} Conceptualization, Methodology, Formal analysis, Software, Writing - Original draft, Writing - Review \& Editing, Visualization 
\textbf{Charlotte Micheloud:} Methodology, Formal analysis, Software, Writing - Original draft, Writing - Review \& Editing, Visualization
\textbf{Stella Erdmann:}  Conceptualization, Methodology, Formal analysis, Writing - Review \& Editing
\textbf{Rachel Heyard:}  Data curation, Writing - Review \& Editing
\textbf{Leonhard Held:} Conceptualization, Methodology, Formal Analysis, Writing - Original draft, Writing - Review \& Editing, Visualization 

\subsection{Acknowledgements}
We thank the RCT DUPLICATE team for their effort and for making their data publicly available. 

\bibliographystyle{unsrtnat}
\bibliography{scepticalP}

\appendix
\counterwithin{figure}{section}
\counterwithin{table}{section}
\section{Replication power}\label{sec:powerApp}
\begin{figure}[!h]
    \centering
   \includegraphics[width = \textwidth]{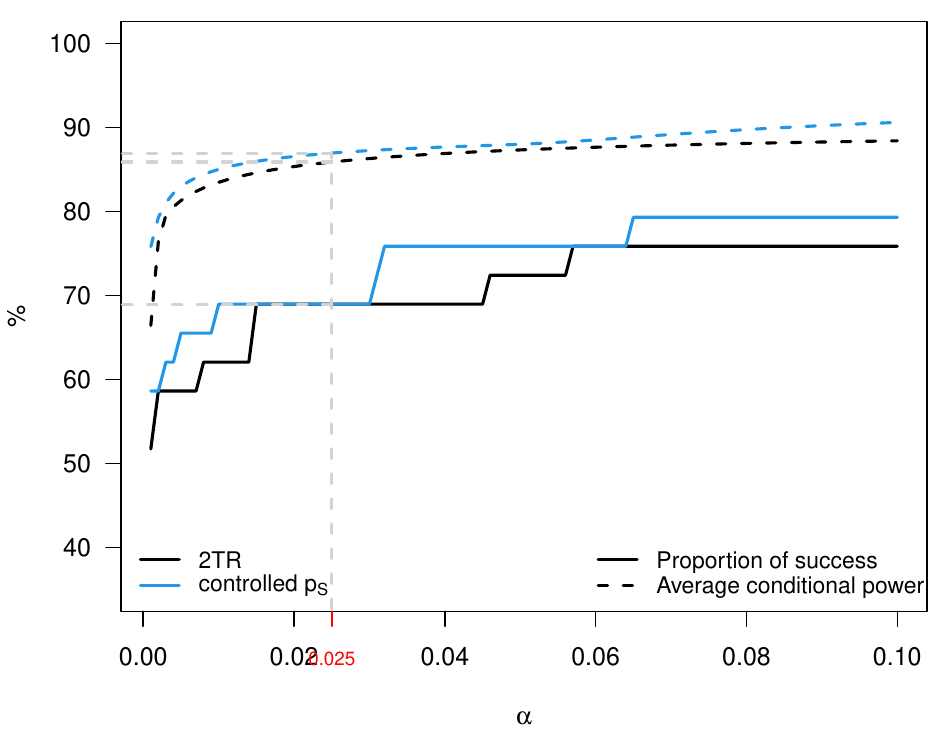}
    \caption{Proportion of success and average predictive power as a function of the one-sided level $\alpha$.}
    \label{fig:propSuccess}
\end{figure}

Figure~\ref{fig:propSuccess} shows the proportion of replication success as well as the average predictive power as a function of the one-sided level $\alpha$.
The proportion of success is for some values of $\alpha$ slightly smaller with the two-trials rule than with the sceptical $p$-value, but the two
methods are in general very close to each other.

\section{Confidence intervals for combined effects}\label{sec:CIApp}
\begin{table}[ht]
\centering \small
\begin{tabular}{rlccccc}
  \hline
& Study & NI & RCT           & pooled RWE   & meta analysis & sceptical $p$-value \\ 
&        & margin & HR (95\% CI)  & HR (95\% CI) & HR (95\% CI)  & (one-sided 97.5\% CI) \\
  \hline
1 & LEADER          & 1.30 & 0.87 (0.78; 0.97) & 0.82 (0.76; 0.87) & 0.83 (0.79; 0.88) & (0; 0.91) \\ 
2 & DECLARE         & 1.30 & 0.83 (0.73; 0.95) & 0.69 (0.59; 0.81) & 0.77 (0.70; 0.85) & (0; 0.88) \\ 
3 & EMPA-REG        & 1.30 & 0.86 (0.74; 0.99) & 0.83 (0.73; 0.95) & 0.84 (0.76; 0.93) & (0; 0.92) \\
4 & CANVAS          & 1.30 & 0.86 (0.75; 0.97) & 0.77 (0.70; 0.85) & 0.80 (0.74; 0.87) & (0; 0.91) \\
5 & CARMELINA       & 1.30 & 1.02 (0.89; 1.17) & 0.90 (0.84; 0.96) & 0.92 (0.87; 0.98) & (0; 1.07) \\
6 & TECOS           & 1.30 & 0.98 (0.88; 1.09) & 0.89 (0.86; 0.91) & 0.90 (0.87; 0.92) & (0; 1.01) \\
7 & SAVOR-TIMI      & 1.30 & 1.00 (0.89; 1.12) & 0.81 (0.76; 0.86) & 0.85 (0.80; 0.90) & (0; 1.04) \\
8 & TRITON-TIMI     & 1.00 & 0.81 (0.73; 0.90) & 0.88 (0.79; 0.97) & 0.84 (0.79; 0.91) & (0; 0.92) \\
9 & PLATO           & 1.00 & 0.84 (0.77; 0.92) & 0.92 (0.83; 1.02) & 0.87 (0.82; 0.93) & (0; 0.96) \\
10 & ARISTOTLE      & 1.38 & 0.79 (0.66; 0.95) & 0.68 (0.61; 0.76) & 0.71 (0.64; 0.78) & (0; 0.85) \\
11 & RE-LY          & 1.46 & 0.66 (0.53; 0.82) & 0.73 (0.60; 0.90) & 0.70 (0.60; 0.81) & (0; 0.80) \\
12 & ROCKET-AF      & 1.46 & 0.79 (0.66; 0.96) & 0.70 (0.62; 0.80) & 0.73 (0.65; 0.81) & (0; 0.85) \\
13 & EINSTEIN-DVT   & 2.00 & 0.68 (0.44; 1.04) & 0.75 (0.62; 0.90) & 0.74 (0.62; 0.88) & (0; 0.86) \\
14 & EINSTEIN-PE    & 2.00 & 1.12 (0.75; 1.64) & 0.67 (0.55; 0.80) & 0.74 (0.62; 0.87) & (0; 1.29) \\
15 & RECOVER II     & 2.75 & 1.08 (0.64; 1.80) & 1.15 (0.74; 1.78) & 1.12 (0.80; 1.57) & (0; 1.48) \\
16 & AMPLIFY        & 1.80 & 0.84 (0.60; 1.18) & 0.81 (0.54; 1.23) & 0.83 (0.64; 1.07) & (0; 1.03) \\
17 & RECORD1        & 1.95 & 0.25 (0.14; 0.47) & 0.17 (0.10; 0.29) & 0.20 (0.13; 0.30) & (0; 0.33) \\
18 & TRANSCEND      & 1.00 & 0.92 (0.81; 1.05) & 0.88 (0.81; 0.96) & 0.89 (0.83; 0.96) & (0; 0.97) \\
19 & ON-TARGET      & 1.13 & 1.01 (0.94; 1.09) & 0.83 (0.77; 0.90) & 0.92 (0.87; 0.97) & (0; 1.04) \\
20 & HORIZON-PIVOTAL& 1.00 & 0.59 (0.42; 0.83) & 0.75 (0.58; 0.97) & 0.69 (0.56; 0.84) & (0; 0.84) \\
21 & DAPA-CKD       & 1.00 & 0.61 (0.51; 0.72) & 0.80 (0.52; 1.26) & 0.63 (0.54; 0.74) & (0; 0.99) \\
22 & PARADIGM-HF    & 1.00 & 0.80 (0.73; 0.87) & 1.02 (0.91; 1.14) & 0.88 (0.82; 0.94) & (0; 1.07) \\
23 & P04334         & 1.00 & 0.56 (0.44; 0.72) & 0.78 (0.62; 0.97) & 0.67 (0.57; 0.79) & (0; 0.86) \\
24 & D5896          & 2.00 & 1.07 (0.70; 1.65) & 1.38 (0.90; 2.13) & 1.21 (0.90; 1.65) & (0; 1.68) \\
25 & IMPACT         & 1.00 & 0.85 (0.80; 0.90) & 1.13 (1.04; 1.23) & 0.93 (0.89; 0.98) & (0; 1.17) \\
26 & POET-COPD      & 1.00 & 0.83 (0.77; 0.90) & 1.02 (0.93; 1.12) & 0.90 (0.85; 0.96) & (0; 1.06) \\
27 & INSPIRE        & 1.00 & 0.97 (0.84; 1.12) & 0.93 (0.90; 0.96) & 0.93 (0.90; 0.96) & (0; 1.01) \\
28 & CAROLINA       & 1.30 & 0.98 (0.84; 1.14) & 0.91 (0.79; 1.05) & 0.94 (0.85; 1.05) & (0; 1.05) \\
29 & PRONOUNCE      & 1.00 & 1.28 (0.59; 2.79) & 1.35 (0.94, 1.93) & 1.34 (0.96; 1.85) & (0; 1.85) \\ 
   \hline
\end{tabular}
\caption{Confidence intervals (CIs) for the hazard ratio (HR) from each RCT and pooled RWE. 
Combined confidence intervals based on fixed effect meta-analysis and sceptical $p$-value are also reported, as well as the non-inferiority (NI) margin (1.00 for superiority studies). 
 } 
\label{tbl:CIs}
\end{table}
\end{document}